**3D Scaffolded Nickel-Tin Li-Ion Anodes with Enhanced Cyclability**

*Huigang Zhang, Tan Shi, Paul V Braun\**

Dr. H. Zhang,
Department of Energy Science and Engineering, Nanjing University, Jiangsu 210093, China
Dr. H. Zhang, T. Shi, Prof. P. V. Braun
Department of Materials Science and Engineering, Materials Research Laboratory, and Beckman Institute, University of Illinois at Urbana-Champaign, Urbana IL 61801, US
E-mail: pbraun@illinois.edu

Abstract: A 3D mechanical stable scaffold is shown to accommodate the volume change of a high specific capacity nickel-tin nanocomposite Li-ion battery anode. When the nickel-tin anode is formed on an electrochemically inactive conductive scaffold with an engineered free volume and controlled characteristic dimensions, it exhibits significantly improved the cyclability.

High energy lithium-ion (Li-ion) batteries are receiving considerable attention because of the increasing importance of high energy density storage technologies. Commercial Li-ion batteries use carbon anodes and open frameworks compounds as cathodes (**Figure 1**a), and considerable efforts are being made to develop higher energy density anode and cathode materials. For example, tin alloy-based anodes are considered promising due to their high volumetric and gravimetric capacities relative to carbon.[1] However, the large volume changes during lithiation and delithiation disintegrates them, leading to rapid capacity loss.[2-4] *In-situ* characterization has identified that microscale pores are formed due to the large tensile stresses during cycling.[5] which lead to macroscale fractures.[2] Once fracture is initiated, the newly exposed surfaces become coated with a solid electrolyte interface (SEI) layer, which consumes cyclable lithium atoms, and may electronically isolate the active material (Figure 1b) causing capacity fade. Recent theoretical and experimental investigations have found that the mechanical stability of Sn and the related Si-based alloy anodes is closely related to the characteristic dimensions (e.g. particle size or film thickness).[6] Large particles or thick films are inclined to fracture, while small particles or thin films are more stable. Nanoscale composite architectures have been used to manage issues related to expansion and fracture in the rather mature Ni/Cd and lead/acid battery systems,[1] and strides have now been made to apply these concepts to high energy density Li-ion chemistries.[7-9] For example, nanocomposites of Ni-Sn and Cu-Sn, where the Ni and Cu serve as an inactive matrix and the Sn as the active material, show greater resistance to damage caused by lithium insertion.[4, 10-13] Framework concepts based on carbon nanotubes,[14] nanorods,[7, 15, 16] and nanocones,[17] metal foams,[10, 18] and metal meshes[19] are also being explored. In general, the goal is to design a structure which accepts at least some of the strain induced by lithium insertion or extraction while maintaining mechanical integrity, and providing a continuous pathway for charge transfer. Because the scaffold needs to provide both mechanical support and electron conduction, metallic or highly conductive carbon scaffolds are generally preferred. The scaffold should not alloy with lithium over the required electrochemical potential range, and the minimum mass and volume of scaffold material possible should be used to maximize the energy density of the electrode.

Because the volume changes in the alloy during cycling are large (usually >100%), there must be space available within the scaffold to accommodate the swelling during lithiation, or the entire electrode will swell, and probably crack.[8] Hollow,[21] ordered macroporous,[4, 11, 22] and reticular [23] tin alloy structures have demonstrated the effectiveness of engineering empty space. However, most of these structures are not scaffolded with inert frameworks, and thus, after cycling, the porous structures become severely deformed.[4] In addition, the tin loading of these structures is generally low relative to that required for a practical anode. Here we show a 3D nickel scaffold supported Ni-Sn alloy structure (Figure 1d) which provides an inactive conductive scaffold, engineered free volume, and efficient ion and electron transport pathways. Within the alloy, Ni buffers the volume expansion as the Sn is reversibly lithiated and delithiated. The overall volume expansion and shrinkage are accommodated by the 3D Ni scaffold. The Ni supported Ni-Sn anodes exhibit both good cycling performance and high energy density. In the initial 5 cycles, a Ni scaffold supported 5.7% Ni and 94.3% Sn nanocomposite anode retains 92% of its capacity, and only loses another 8% over the next 195 cycles. The volumetric capacity of this electrode is 1846 mAh cm$^{-3}$, which is 90% of the theoretical volumetric capacity of lithiated tin (~2106 mAh cm$^{-3}$), and about twice that of carbon (837 mAh cm$^{-3}$). The Sn content in the alloy can be engineered by varying the composition of the plating bath providing opportunities for optimization.

The electrode is fabricated by first depositing uniform polystyrene (PS) spheres on a tungsten substrate by slow evaporation of water from a colloidal suspension followed by annealing near the glass transition temperature of PS to strengthen the opal and enlarge the contact points between spheres. Nickel is plated through the porous opal using a commercial electroplating solution. The PS spheres are removed by tetrahydrofuran (THF) yielding the inverse opal template (**Figure 2**a). Finally, the porosity of the nickel inverse opal is increased by electropolishing. The resulting highly porous structure is shown in Figure 2b. For complete nickel scaffold fabrication details, refer to our published procedures.[24, 25] The electrolytically active Ni-Sn alloy is deposited from a pyrophosphate-containing plating solution (**Table 1**). The Ni/Sn ratio of deposit, which is controlled by the solution composition, is quantified by inductively coupled plasma atomic emission spectroscopy (ICP-AES). Table 1 shows three typical compositions. Galvanostatic pulses are used for electrodeposition, and the number of pulses determines the Ni-Sn film thickness. Figure 2c shows a scanning electron microscopy (SEM) image of the alloy-coated nickel inverse opal.

The Ni scaffold is not completely filled with Ni-Sn to provide space for the alloy anode materials to expand inside spherical void during lithiation. Knowing the volume change upon lithiation, the preferred thickness of the plated alloy layer and the porosity of nickel scaffold can be optimized by geometric analysis.[24, 26] After 200 lithiation and delithiation cycles, the alloy electrode remains flat without delamination, indicating that the electrode integrity is maintained after multiple volume changes. Because of the residual electrolyte and SEI layers, the periodic structure is not observed when the electrode is directly imaged (Figure 2d) (SEM images of a cleaned sample where the structure is apparent are discussed later, and shown in **Figure 6**).

**Figure 3**a shows X-ray diffraction (XRD) patterns of the three alloys (S1, S2, and S3) prepared from the plating solutions listed in Table 1. XRD of S1 shows a metastable phase, which contains 66.9% Sn (Ni:Sn 1:1 molar ratio). In S3, a pure tin phase appears along with a metastable phase. S2 is identified as $Ni_3Sn_4$. The results are in agreement with a previous report.[27] The appearance of metastable Ni-Sn alloy phases has been reported,[27-29] and the formation mechanism and analysis of electrodeposited Ni-Sn metastable phases can also be found in the literature.[30] Generally, the tin content in the deposit increases as the tin concentration in the plating solution increases.

It is known that Ni-Sn alloy electrodes have a first cycle activation step followed by a reversible electrochemical process: $Sn + xLi \leftrightarrow Li_xSn$.[16, 31] To characterize the tin lithiation/delithiation, cyclic voltammetry (CV) (Figure 3b) was conducted at a scan rate of 0.05 mV s$^{-1}$ for S1, S2, and S3 after two galvanostatic charge/discharge cycles. In the anodic scan of three samples, several peaks appear at ~0.49 and 0.58~0.61 V.vs Li/Li$^+$. The cathodic scans show peaks at ~0.32 and 0.45 V. Each peak represents the electrochemical reaction between two individual lithium-tin alloy phases and lithium. Although it is hard to assign each peak to two distinct phases [32] because scanning rates and crystallite sizes can complicate the interpretation of CV peaks, the peaks between 0.8 and 0.1V are typical characteristics of multistep lithiation/delithiation for tin,[33] in agreement with previous reports.[12, 28] Multiple alloying steps are further confirmed by the ultra-slow constant current cycling data in **Figure 4**b.

To identify if the 3D scaffolded electrode can combine the stabilizing mechanisms described in the first paragraphs, 3D scaffolded and flat Ni-Sn (as shown in Figure 4a) are assembled into coin cells with lithium counter electrodes, respectively. The S3 composition in Table 1 is used for Sn deposition. The areal capacity of loaded Ni-Sn is controlled to be around 1.2 mAh cm$^{-2}$ for both the 3D and flat samples. The charge/discharge curves in Figure 4b shows that both 3D and flat samples have similar voltage profiles with multiple plateaus, which is in agreement with the CV results. The gap of about 0.2 V between charge and discharge is due to rate-dependent polarization and thermodynamic hysteresis.[2, 34] Flat Ni-Sn almost shows no capacity above 0.7 V but 3D scaffolded Ni-Sn has ~4% capacity between 2.0 and 0.7 V during lithiation. Compared to the flat plateau around 0.4V on the lithiation curve of the Ni-Sn film, the 3D scaffolded Ni-Sn shows a slightly sloped plateau. Capacity retention at various rates is shown in Figure 4c. At 0.7C, flat Ni-Sn film loses 19% of its 0.5C capacity. Its capacity retention decreases sharply from 72% at 3C to 2.8% at 7.6C. 3D scaffolded Ni-Sn demonstrates the capacity retention of 78% at 5C and 35% at 30C. Even at 50C, 3D scaffolded Ni-Sn is able to deliver 8% of its 0.5C capacity. The enhanced rate properties could be attributed to the much efficient electron and ion pathways that 3D scaffold provides, which is in accordance with the electrochemical impedance spectroscopy (EIS) measurements. Figure 4d shows that 3D scaffolded Ni-Sn has much lower impedance than flat film. Note that the semicircle in the Nyquist plot is depressed, indicating the overlap of multiple semicircles (each semicircle embodies one interface on the way of electrons or/and ions and is usually modeled as the parallel combination of a capacitor and a resistor). The semicircle in the high frequency range is correlated to the resistance of the SEI film.[35]

S1, S2, and S3 are galvanostatically cycled 200 times between 0.02 and 1.5V using a lithium counter electrode (**Figure 5**). Most tin-based anodes have high initial capacities which fade quickly as the electrode and/or the SEI cracks. Whenever fresh electrode surface is exposed, additional SEI is formed, which can be observed as a low coulombic efficiency (CE). For S3, the first cycle CE is 91.6%, which first increases rapidly, then gradually increases until it is 99.7% after 200 cycles (Figure 5). The high CE is a strong indication that the SEI is stable with cycling, even with the large volume changes in the electrode.

S1, S2, and S3 show first cycle capacities of 597, 729, and 895 mAh $g^{-1}$ (active materials basis), respectively, which decrease most significantly over the first few cycles (Figure 5). For example, S3 (94.3% Sn) loses 8% of its initial capacity over the first 5 cycles and another 8% capacity over the next 195 cycles. The flat film sample shows an initial capacity of about 710 mAh $g^{-1}$ which first decreases to 589 mAh $g^{-1}$ and then increases to 639 mAh $g^{-1}$ by the sixth cycle. The decrease followed by an increase may be because the cycling rate is too high to fully access all the active materials until the electrode starts to crack. After 20 cycles, the capacity abruptly decreases to several tens of mAh $g^{-1}$. The large oscillation of CE around 100% implies a process in which the active materials may electrically connect or disconnected due to volume changes during cycling. Such a large CE change could not be explained only by SEI formation which is accompanied only by CEs <100%. The relatively stable CE of three scaffolded samples indicates that 3D scaffold could accommodate the large volume change. To further explore the dynamic processes during cycling, a control experiment is designed as shown in Figure 6. A Ni-Sn alloy is electrodeposited on both the 3D structured and flat regions of one sample (Figure 6c). Both regions are cycled under the same electrochemical conditions. After the sample is fully delithiated, it is washed with fresh dimethylene carbonate (DMC). Any residue insoluble in DMC is then removed with water. It can be clearly seen in Figure 6a that although the electrode surface becomes rough due to repeated alloying and de-alloying, the three-dimensionally structured electrode maintains its integrity. In the flat region (Figure 6b), the alloy film is cracked and partially delaminated, indicating the deformation and fracture in flat films [4] can be managed using a three-dimensionally structured metal scaffold under the same electrochemical conditions. Importantly, the scaffold enables nearly complete decoupling of the total energy density and the thickness of the active layer. The electrochemically active layer appears to be stabilized by the scaffold enabling a system with a high energy density without the problems associated with traditional thicker alloy electrodes.

In conclusion, enhanced cyclability of a Ni-Sn alloy electrode is demonstrated using a designed electrically conductive 3D scaffold. The scaffolded composite alloy electrode shows robust mechanical stability and improved cyclability, presumably because the scaffold can accommodate the volume changes and maintain electric connections during lithiation and delithiation. The method described is general, and has potential to be applied to other high capacity anode materials including Si, Sb, Al, $SnO_2$, $Fe_3O_4$, and $Co_3O_4$. Most importantly, this approach enables coupling the cycling stability of a thin film with the high capacity of thick electrode.

**Experimental**

A tungsten foil (Sigma-Aldrich Corp) is used for the opal deposition. The foil first is annealed in forming gas (5% $H_2$ and 95% Argon) at 550 °C for 2.5 h. PS spheres purchased from Invitrogen Corp. are diluted to 0.2% solid content. The annealed foil is vertically placed into a small vial containing the PS suspension at 55 °C. When about 95vol% of the PS suspension is evaporated, the sample is removed from the vial and dried at room temperature. The samples are then annealed at 95 °C for 2 h. Using a nickel plate counter electrode, nickel is electroplated into the 3D PS structures at a current density of ~1.5 mA $cm^{-2}$ using SN-10 nickel plating solution (Transene). After Ni electrodeposition is complete, the sample is rinsed with de-ionized water and dried. THF is used to remove the PS template. The electropolishing solution containing 0.5~1 M $Ni^{2+}$ is prepared by electrochemically dissolving Ni metal into EP1250 (Technic Corp) at 6 V. The macroporous nickel is electropolished in this solution by a series of 6 V pulses (80 ms on and 16 s off) at 60°C. The extent of polishing is controlled by the number of pulses (usually ~60). Detailed experimental procedures for the macroporous nickel preparation can be found in our previous publications.[24, 25, 26]

The Ni-Sn alloy is electrodeposited into the Ni scaffold using the plating solutions in Table 1. A pure Sn electrode is used as the counter electrode. A series of galvanostatic pulses are applied (2 mA $cm^{-2}$ for 2 s and 0 mA $cm^{-2}$ for 7 s) between the macroporous nickel samples and the counter electrode. Each Ni-Sn composition in Table 1 is also electrodeposited on ITO coated glass and the resulting metal deposit is transferred to Scotch tape for x-ray diffraction (Philips X'pert MRD) and ICP-AES (Optima 2000 DV, Perkin Elmer) analysis. The electrode morphology is observed using a Hitachi 4800 SEM. Electrochemical analysis is conducted using Princeton Applied Research Model 273A and Bio-logic VMP3 potentiostats. The Ni-Sn anode is cycled using a lithium foil as the counter electrode. EIS is measured with lithium as the reference and counter electrode in a three-electrode cell (ECC-REF, Bio-logic Inc.).The electrolyte is 1 M LiClO$_4$ in 1:1 mass ratio mixture of ethylene carbonate and dimethylene carbonate.

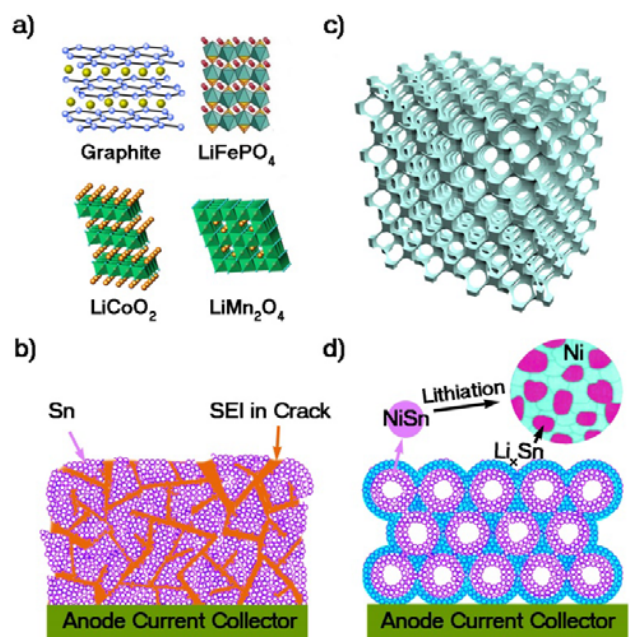

**Figure 1.** a) Anode (graphite) and cathode intercalation materials, which have good electrochemical cyclability due to their stable frameworks. b) A pure metal anode, which pulverizes during cycling. c) 3D electrochemically inactive electrode scaffold. d) Composite alloy anode consisting of a 3D nickel porous scaffold coated with Ni-Sn (when the Ni-Sn alloy is charged, only Sn is lithiated and Ni acts as an inactive matrix which buffers the expansion of $Li_xSn$) as shown in the "Mosaic" domain separation resulting from lithiation (drawing modified and redrawn from Ref [20]).

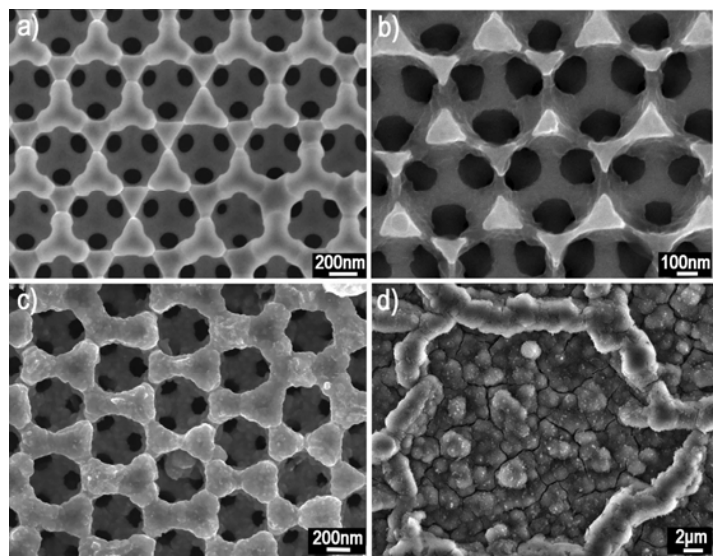

**Figure 2**. SEM images of a) the Ni inverse opal, b) the electropolished Ni scaffold, c) 3D Ni scaffolded Ni-Sn electrode, and d) the alloy electrode after 200 cycles (note, in d, the scale bar is significantly larger than in a-c).

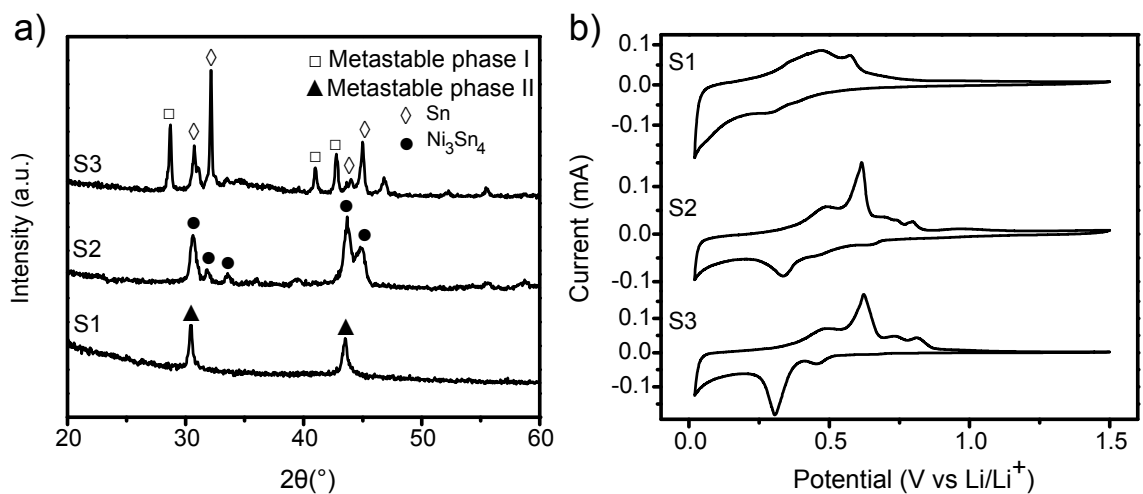

**Figure 3**. a) XRD and b) Cyclic voltammetry of samples S1, S2, and S3 (scan rate 0.05mV s$^{-1}$).

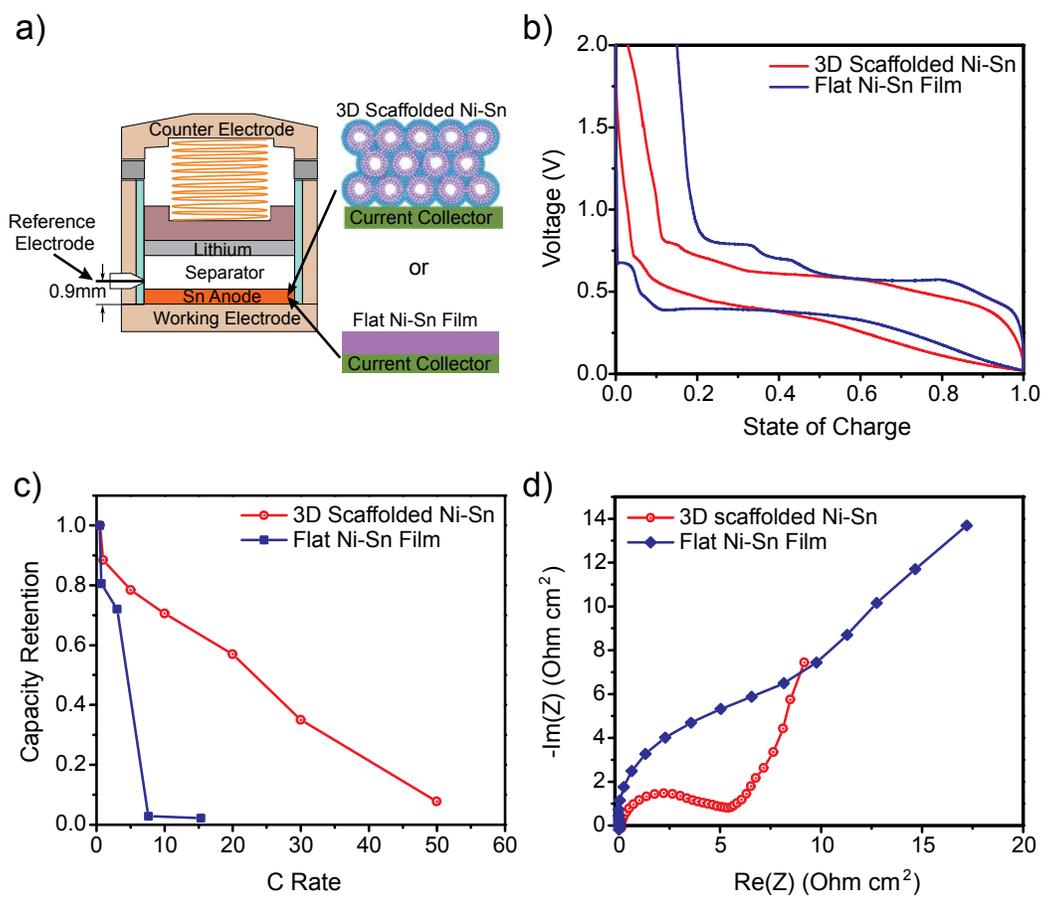

**Figure 4**. a) Illustrations of the 3D scaffold Ni-Sn sample, a flat Ni-Sn sample, and the 3-electrode test cell. b) Charge/discharge curves of the scaffolded and flat samples at ~0.05C. c) Capacity retention of scaffolded and flat samples at various C rates. d) EIS of the scaffolded and flat samples vs Li using the 3-electrode cell shown in a). Before measurements, the cells are equilibrated at 0.2 V for 1 h.

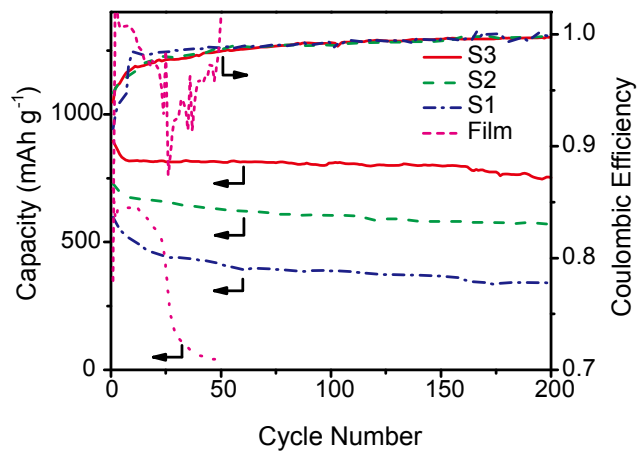

**Figure 5**. Capacity and CE (active materials basis) of S1, S2, S3, and flat Ni-Sn samples cycled at 0.5C between 0.02 and 1.5V vs. a lithium counter electrode.

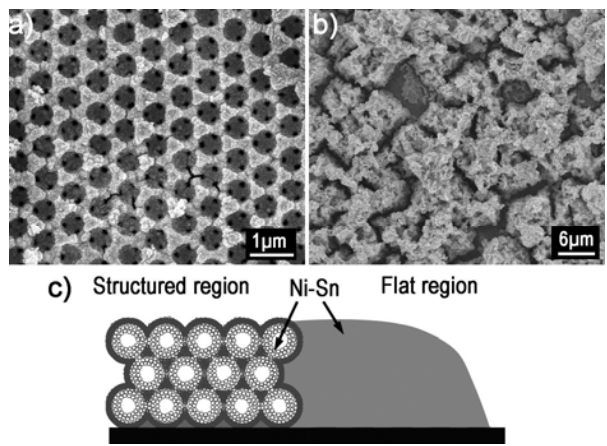

**Figure 6.** SEM images of a Ni-Sn anode after cycling. a) a structured porous nickel scaffold region, and b) a flat region. c) a cross-sectional schematic.

**Table 1.** Ni-Sn alloy electroplating solutions used to grow samples S1, S2, and S3 and the measured wt% Sn of each sample.

| Solution Component | Composition (g/L) | | |
| --- | --- | --- | --- |
| | S1 (66.9 wt% Sn) | S2 (78.9wt% Sn) | S3 (94.3 wt% Sn) |
| $NiCl_2$ | 12 | 8 | 8 |
| $SnCl_2 \cdot 2H_2O$ | 10 | 20 | 46 |
| $K_4P_2O_7$ | 300 | 300 | 300 |
| Glycine | 8 | 8 | 8 |
| Potassium sodium tartrate | 8 | 8 | 8 |